# ABELIAN ACTION FOR QUARK CONFINEMENT: A DIRECT EVALUATION

KEN YEE*

*Physics and Astronomy, LSU, Baton Rouge, LA 70803-4001, USA*
kyee@rouge.phys.lsu.edu

## ABSTRACT

We evaluate $S_{APQCD}$, the Abelian projection QCD(APQCD) action, using the microcanonical demon method. For $SU(2)$, we find that $S_{APQCD}$ at strong coupling is essentially the compact QED(CQED) action with $\beta_{CQED} = \frac{1}{2}\beta_{SU(2)}$. Since CQED deconfines when $\beta_{CQED} > 1$, this relation must break down as $\beta_{SU(2)} \to 2$. Indeed we find $S_{APQCD}$ mutates: near $\beta_{SU(2)} \sim 2$ it gains additional operators, including an exogenous *negative* magnetic monopole mass shift. Since monopoles are condensed in CQED when $\beta_{CQED} < 1$, a vicarious corollary of these results is that $SU(2)$ monopoles are condensed when $\beta_{SU(2)} < 2$. $S_{APQCD}$ for $SU(3)$ has similar behavior.

A clear demonstration that monopole condensation is the origin of QCD confinement would be a notable achievement. To this end, 't Hooft[1-3] proposed that QCD monopoles are magnetic with respect to the $[U(1)]^{N-1}$ Cartan subgroup of color $SU(N)$. Full $SU(N)$ gauge symmetry obscures these charges and it is necessary to gauge fix at least the $SU(N)/[U(1)]^{N-1}$ symmetry to expose them. In this scenario monopoles are fixed-gauge manifestations of gauge field features responsible for QCD confinement. Only in special gauges does one have a picture of QCD confinement caused by monopole condensation. In other gauges the gauge field features causing confinement are still present but they do not look like magnetic monopoles.[4]

Numerical studies have found that maximal Abelian(MA) gauge[5] is compelling for 't Hooft's hypothesis. Upon decomposing gauge field $A$ into purely diagonal($n$) and purely off-diagonal($ch$) parts

$$A = A^n + A^{ch}, \qquad (1)$$

the MA gauge condition $D_\mu^n A_\mu^{ch} \equiv \partial_\mu A_\mu^{ch} - ig[A_\mu^n, A_\mu^{ch}] = 0$ leaves a residual $[U(1)]^{N-1}$ gauge invariance under $\Omega_{\text{residual}} = \text{diag}(\exp^{-i\omega_1}, \cdots, \exp^{-i\omega_N})$ where $\sum_{i=1}^N \omega_i = 0$. Under $\Omega_{\text{residual}}$ the $N$ diagonal matrix elements $(A^n)_{ii}$ transform as neutral photon fields whereas the $N(N-1)$ offdiagonal matrix elements $(A^{ch})_{ij}$ transform as charged matter fields: $(A_\mu^n)_{ii} \to (A_\mu^n)_{ii} - \frac{1}{g}\partial_\mu \omega_i$ and, for $i \neq j$, $(A_\mu^{ch})_{ij} \to (A_\mu^{ch})_{ij} \exp^{-i(\omega_i - \omega_j)}$. Since

---

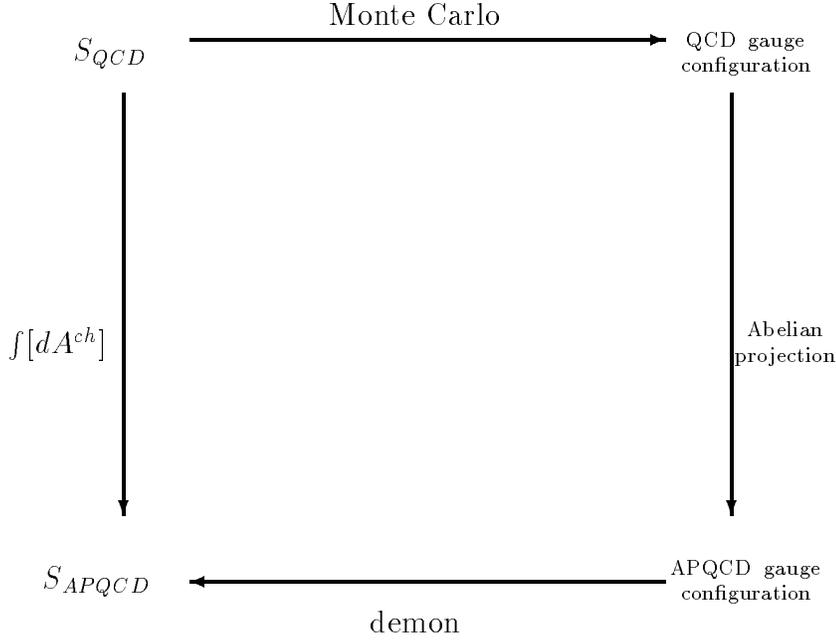

Fig. 1. To integrate out $A^{ch}$ we: (i)generate an importance sampling $SU(N)$ gauge configuration; (ii)project this configuration to a $[U(1)]^{N-1}$ configuration; and (iii)compute the $S_{APQCD}$ couplings using the microcanonical demon.

$(A^{ch})_{ij}$ carries two different $U(1)$ charges, the $A^{ch}$ fields induce "interspecies" interactions between the $N$ photons. On the lattice the monopole currents are identified according to discretized versions[6] of $k_\mu \equiv \frac{1}{2\pi}\epsilon_{\mu\nu\lambda\delta}\partial_\nu f_{\lambda\delta}$ and $f_{\mu\nu} \equiv \partial_\mu A^n_\nu - \partial_\nu A^n_\mu$.

This procedure where only the diagonal $A^n$ components of the $SU(N)$ gauge fields are used for measuring $k_\mu$ and $f_{\mu\nu}$ is called *Abelian projection*. Since $\sum_{i=1}^{N}(A^n_\mu)_{ii}$ is invariant under $\Omega_{\text{residual}}$, an irreducible representation of $[U(1)]^{N-1}$ is $\theta^i_\mu \equiv (A^n_\mu)_{ii} - \Lambda_\mu$ where $\Lambda_\mu \equiv \frac{1}{N}\sum_{j=1}^{N}(A^n_\mu)_{jj}$. While vector field $\Lambda$ is $[U(1)]^{N-1}$ invariant, the $\theta^i$ transform as $\theta^i_\mu \to \theta^i_\mu - \frac{1}{g}\partial_\mu \omega_i$ and obey constraint $\sum_{i=1}^{N}\theta^i_\mu = 0$. We shall refer to the quantum dynamics of the $N$ angles $\theta^i$ as Abelian projected QCD or APQCD. Equivalently, APQCD is the field theory obtained by integrating out $A^{ch}$ and $\Lambda$ from QCD in MA gauge.[7] Its action $S_{APQCD}$ is formally defined as

$$-S_{APQCD}[\theta^1,\cdots,\theta^N] \equiv \log\left\{\int [dA^{ch}d\Lambda]\ \exp(-S_{QCD})\ \Delta_{FP}\ \delta[D^n_\mu A^{ch}_\mu]\right\}. \qquad (2)$$

Monopoles arise in APQCD due to topological quantum fluctuations in the compact fields $\theta^i$.

While there is no guarantee that $S_{APQCD}$ has a simple form or is otherwise well-behaved, it is of central import due to *Abelian dominance*,[8] the fact that $\theta^i$ Wilson loops in APQCD have predominantly the same string tension as $SU(N)$ Wilson loops in QCD. Abelian dominance has the following formal implication. If tr**W** is an $SU(N)$ Wilson loop and if $\langle\cdot\rangle_{QCD}$ and $\langle\cdot\rangle_{APQCD}$ refer respectively to $S_{QCD}$ and $S_{APQCD}$

expectation values, the APQCD operator $\mathcal{W}$ which obeys

$$\langle \mathcal{W} \rangle_{APQCD} = \text{tr}\langle \mathbf{W} \rangle_{QCD} \tag{3}$$

is

$$\mathcal{W} = \exp(+S_{APQCD}) \int [dA^{ch} d\Lambda] \, \exp(-S_{QCD}) \, \Delta_{FP} \, \delta[D_\mu^n A_\mu^{ch}] \, \text{tr}\mathbf{W}. \tag{4}$$

Abelian dominance means that the complicated operator $\mathcal{W}$, which in other gauges would be a superposition of assorted $[U(1)]^{N-1}$-invariant operators of various sizes and shapes, is (for string tension) well-approximated by a $\theta^i$ loop of the *same size and shape* as tr$\mathbf{W}$ in MA gauge. In other gauges the $\langle \mathcal{W} \rangle_{APQCD}$ string tension would be due to a *combination* of $S_{APQCD}$ effects *and* properties of $\mathcal{W}$. An extreme example in which $S_{APQCD}$ is immaterial is in a (hypothetical) gauge where $\mathcal{W} = \exp\{-\lambda RT\} \cdot 1$—the constant APQCD operator with area law coefficient. In such a gauge, the coefficient in $\mathcal{W}$ hordes the area law and $S_{APQCD}$ is immaterial because $\mathcal{W}$ is simply 1. In stark contrast, $S_{APQCD}$ alone determines string tension in MA gauge: given $S_{APQCD}$ one can reconstruct the QCD string tension using APQCD Wilson loops without reference to the RHS of (4). In MA gauge $S_{APQCD}$ apparently knows all about QCD confinement properties.

Our numerical procedure for evaluating $S_{APQCD}$ is depicted in Figure 1. Let us temporarily focus on $SU(2)$. First, we make a representative importance sampling APQCD gauge configuration by applying the Abelian projection to a Monte Carlo lattice $SU(2)$ gauge configuration at some chosen coupling $\beta_{SU(2)}$. Seeking the action $S_{APQCD}$ which would reproduce this APQCD configuration[7] in a Monte Carlo simulation, we state an ansatz for $S_{APQCD}$ and apply the microcanonical demon technique[9] to compute the parameters of this ansatz. This "inverse Monte Carlo" procedure is repeated to determine how $S_{APQCD}$ fluctuates between different importance sampling $SU(2)$ configurations.

The general $U(1)$-invariant action consistent with APQCD involves an infinity of operators. Fortunately, previous studies[7,10] and, independently, the demon technique indicate that neither extended nor highly charged Wilson loops contribute substantially to $S_{APQCD}$. In particular, we have applied the demon to an ansatz consisting of $1 \times 1$, $2 \times 2$, and $3 \times 3$ plaquettes. Over a wide range of $\beta_{SU(2)}$, we are unable to resolve a nonzero signal for any of the $2 \times 2$ or $3 \times 3$ plaquette couplings. Therefore, we focus now on a $1 \times 1$ ansatz

$$-S^{\text{ansatz}} = \sum_{q=1}^{3} \sum_{x,\mu<\nu} \beta_q \cos q\Theta_{\mu\nu} - \kappa \sum_{x,\mu} k_\mu(x) k_\mu(x). \tag{5}$$

$\cos q\Theta_{\mu\nu}$ is a $1 \times 1$ plaquette in $U(1)$ representation $q$ given in terms of link angles $\theta_\mu^1$. $\kappa$ shifts the $q = 1$, $1^3$ monopole mass[11] implicit in $\beta_1 \cos \Theta_{\mu\nu}$, allowing the APQCD monopole mass to be independent of $\beta_1$. Of course, $\beta_i$ and $\kappa$ vary with $\beta_{SU(2)}$.

In our version of the demon technique, imagine a battalion of demons each carrying $M$ *coupled* thermometers corresponding to the $M$ undetermined coefficients in the ansatz action. ($M = 4$ for $S^{\text{ansatz}}$.) The demons thermalize with the APQCD configuration by hopping from link to link and exchanging energy($\equiv$action) with the configuration.[†] The thermometers are coupled by requiring all of their energies to remain within a given range $[-E^0, E^0]$; if any proposed energy exchange violates this range it is rejected. Upon thermalization the couplings are read off from the battalion of energies, which has a Boltzmann distribution. Statistical errors are computed by jackknifing the demons. (The independent errors from jackknifing $SU(2)$ configurations are comparable in size.) In principle, if $S^{\text{ansatz}}$ contains all the operators of $S_{APQCD}$ the demon technique yields all the coupling constants exactly (modulo statistics). In practice, $S^{\text{ansatz}}$ is a truncated action which is unlikely to contain all $S_{APQCD}$ operators. Extensive numerical experiments with idealized configurations[12] reveal that if operators are missing the method yields effective values for the couplings. These effective values would not be the same as the "true" values when all operators are present.

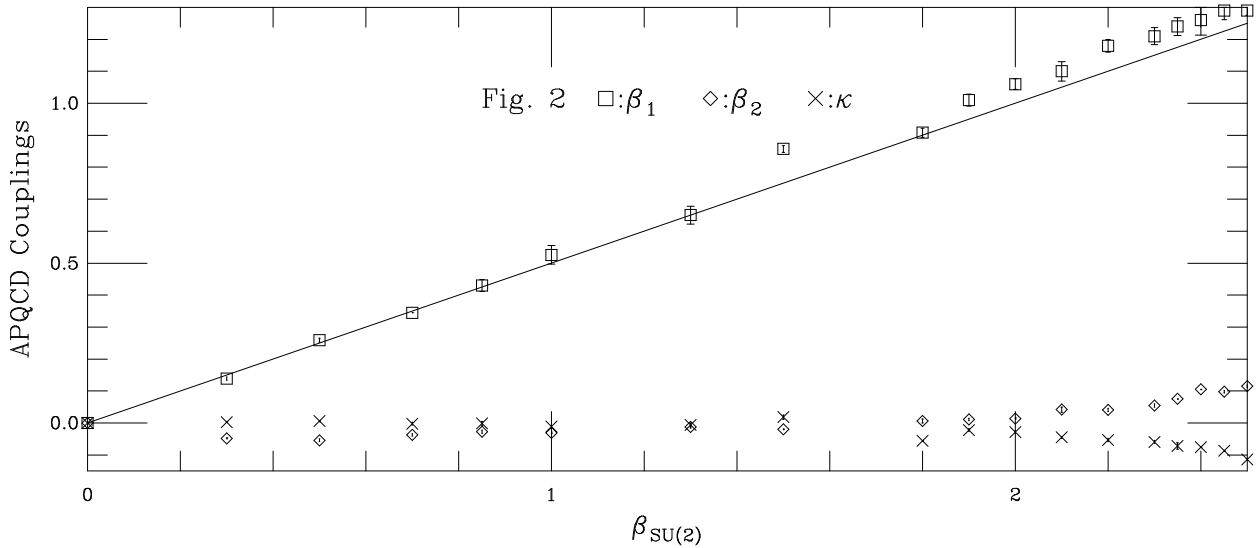

Fig. 2. Figure 2 depicts $S^{\text{ansatz}}$ coefficients $\beta_1$, $\beta_2$ and $\kappa$ as a function of $\beta_{SU(2)}$. $|\beta_3|$, not depicted, is always smaller than $|\beta_2|$, typically by a factor of $3-5$. Our $20^3 \times 16$ lattices are all well inside the zero temperature phase for the $\beta_{SU(2)}$ range depicted. The bold $\beta_1 = \frac{1}{2}\beta_{SU(2)}$ line is a guide-to-eye.

Figure 2 shows $S^{\text{ansatz}}$ coefficients $\beta_1$, $\beta_2$ and $\kappa$, computed by the demon, as a function of $\beta_{SU(2)}$. $|\beta_3|$, not depicted, is always smaller than $|\beta_2|$, typically by a factor of $3-5$. Each $\beta_{SU(2)}$ configuration is generated fresh from a cold start so our data points do not contain any spurious correlations. Our $N_S^3 \times N_T = 20^3 \times 16$ lattices are all well inside the zero temperature phase; for the range of $\beta_{SU(2)}$ shown the APQCD Polyakov loop vanishes. As depicted, at strong coupling($\beta_{SU(2)} < 2$)

$$\beta_1 \sim \frac{1}{2}\beta_{SU(2)}, \quad \beta_{2,3} \sim 0, \quad \kappa \sim 0, \tag{6}$$

that is, $S_{APQCD}$ reduces to the compact QED(CQED) action at strong coupling. At weaker coupling($\beta_{SU(2)} > 2$) $\beta_2$ and $\kappa$ grow in magnitude but $\beta_1$ always remains the largest coupling.

---

[†]Contrary to Ref. 9 we do not update the APQCD configuration so that energy is figuratively rather than literally "exchanged." This shortens the numerical algorithm and avoids the possibility of damaging the APQCD configuration if the battalion of demons absorbs too much energy.

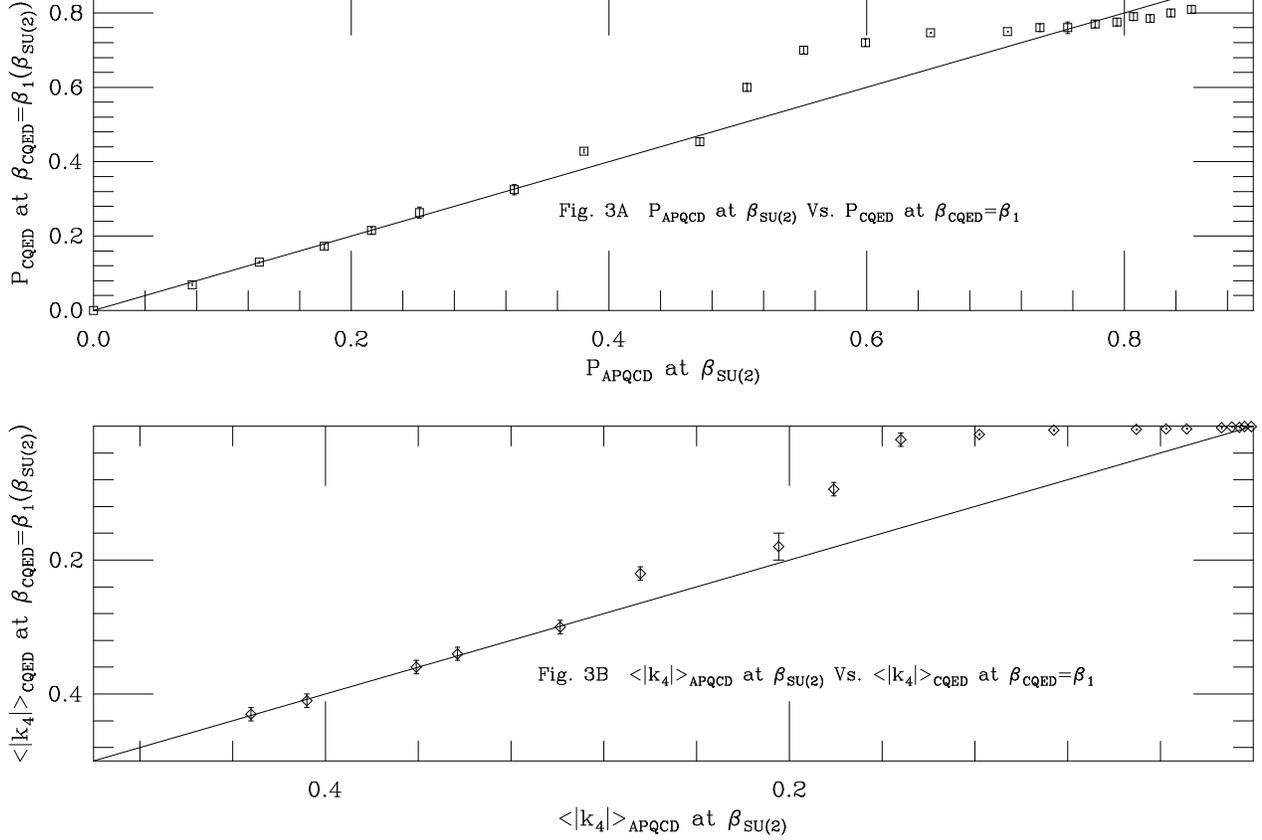

Fig. 3. 3A compares APQCD plaquettes $P_{APQCD}$ at $\beta_{SU(2)}$ to CQED plaquettes $P_{CQED}$ at $\beta_{CQED} = \beta_1(\beta_{SU(2)})$ for a range of $\beta_{SU(2)}$ values. When $\beta_{SU(2)} < 2$ the data points lie on the bold $P_{CQED} = P_{APQCD}$ line showing that $S_{CQED}$ is a good model of $S_{APQCD}$. The set of points wandering off of the $P_{CQED} = P_{APQCD}$ line corresponds to $\beta_{SU(2)} > 2$, when $S_{CQED}$ is not a good model of $S_{APQCD}$. 3B is an analogous plot using monopole densities.

Note that since monopoles are condensed when $\beta_{CQED} < 1$ in[13] CQED, Figure 2 or Eq. (6) proves (albeit vicariously) that $SU(2)$ monopoles are condensed when $\beta_{SU(2)} < 2$.

When $\beta_{SU(2)} > 2$, the situation is not so clear. In fact, Figure 2 suggests a paradox in the $\beta_{SU(2)} > 2$ region: how can APQCD maintain confinement in the continuum limit if CQED deconfines when $\beta_{CQED} > 1$? Clearly, either the meaning or validity of relation (6) must break down when $\beta_{SU(2)}$ is sufficiently large. Either (I)Abelian dominance does not survive the $\beta_{SU(2)} \sim 2$ crossover making $S_{APQCD}$ less pertinent at weaker coupling—see discussion pertaining to Eq. (4); or (II)$S_{APQCD}$ gains additional operators near $\beta_{SU(2)} \sim 2$; or a combination of (I) and (II). We do not have anything to say about (I) in this Note except to observe that there has been no definitive study of Abelian dominance at larger $\beta_{SU(2)}$ values.[‡]

(II) requires that $S_{APQCD}$ is not well described by $S_{CQED}$ when $\beta_{SU(2)} > 2$. Indeed,

---

[‡]In typical string tension measurements, the static quark potential is determined by superimposing data points from a range of different $\beta_{SU(2)}$ values.

we can demonstrate this by simulating

$$-S_{CQED} = \sum_{x,\mu<\nu} \beta_{CQED} \cos\Theta_{\mu\nu} \Big|_{\beta_{CQED}=\beta_1(\beta_{SU(2)})} \quad (7)$$

(also on a $20^3 \times 16$ lattice) to see if it reproduces corresponding APQCD expectation values. As depicted in Figure 3, $S_{CQED}$ reproduces APQCD plaquette averages and monopole densities only in the $SU(2)$ strong coupling region. At weaker coupling the CQED simulations start to disagree dramatically with APQCD. This implies that at weaker coupling either other terms of $S^{\text{ansatz}}$ have become important or $S^{\text{ansatz}}$ itself is inadequate. In any case, this means $S_{APQCD}$ is not form-invariant between the strong and weak coupling regimes: at strong coupling $S_{APQCD}$ is well approximated by $S_{CQED}$; at crossover region $\beta_{SU(2)} \sim 2$ $S_{APQCD}$ mutates and develops substantial deviations from $S_{CQED}$. Inspection of Figure 2 reveals that a possible scenario might be that $\kappa$, the exogenous magnetic monopole mass shift, becomes more and more *negative* at larger $\beta_{SU(2)}$. As negative monopole mass favors monopole condensation (compensating for a large $\beta_1$), the occurrence of a sufficiently negative $\kappa$ in $S_{APQCD}$ at $\beta_{SU(2)} > 2$ could maintain APQCD confinement.

Note that Figure 2, as characterized by Eq. (6), "explains" Abelian dominance—at least in the strong coupling regime. The $SU(2)$ plaquette in the strong coupling expansion behaves like $P_{QCD} \sim \frac{1}{4}\beta_{SU(2)}$ and the CQED plaquette like $P_{CQED} \sim \frac{1}{2}\beta_{CQED}$. Therefore, identifying $P_{CQED}(\beta_{CQED} = \beta_1)$ with $P_{APQCD}$ and applying Eq. (6) yields

$$P_{APQCD} \sim \frac{1}{4}\beta_{SU(2)} \sim P_{QCD}. \quad (8)$$

Carrying this argument over to larger Wilson loops leads to a (strong) statement of Abelian dominance: at sufficiently strong coupling APQCD and QCD Wilson loop averages and, hence, string tensions are equal. Figure 4 confirms (8) and shows how it breaks down at weaker coupling. Note Eq. (8) contradicts the naive expectation, based on $P_{QCD}$ containing a trace over a $2 \times 2$ matrix and $P_{APQCD}$ involving no trace, that $P_{APQCD} = \frac{1}{2}P_{QCD}$.

We have obtained similar results for the $SU(3)$ Abelian projection which will be described elsewhere. For $SU(3)$, $S_{APQCD}$ is more complicated due to interspecies dynamics.[10,14,15] Nonetheless, we have observed the same crossover behavior in $SU(3)$. At strong coupling, $S_{APQCD}$ is simple; at weaker coupling, there is a clear crossover to a more complicated action. A preliminary indication of this can be seen in the data reported in Ref. 3.

In conclusion, $S_{APQCD}$ is not form-invariant between the strong and weak coupling regimes. Thus, glamorous dynamical and phenomenological features of the Abelian projection confinement mechanism, such as Abelian dominance or the issue of whether APQCD is a Type I or Type II superconductor,[16] might vary with lattice spacing. As it is, it is imperative to distinguish between strong and weak coupling data in order

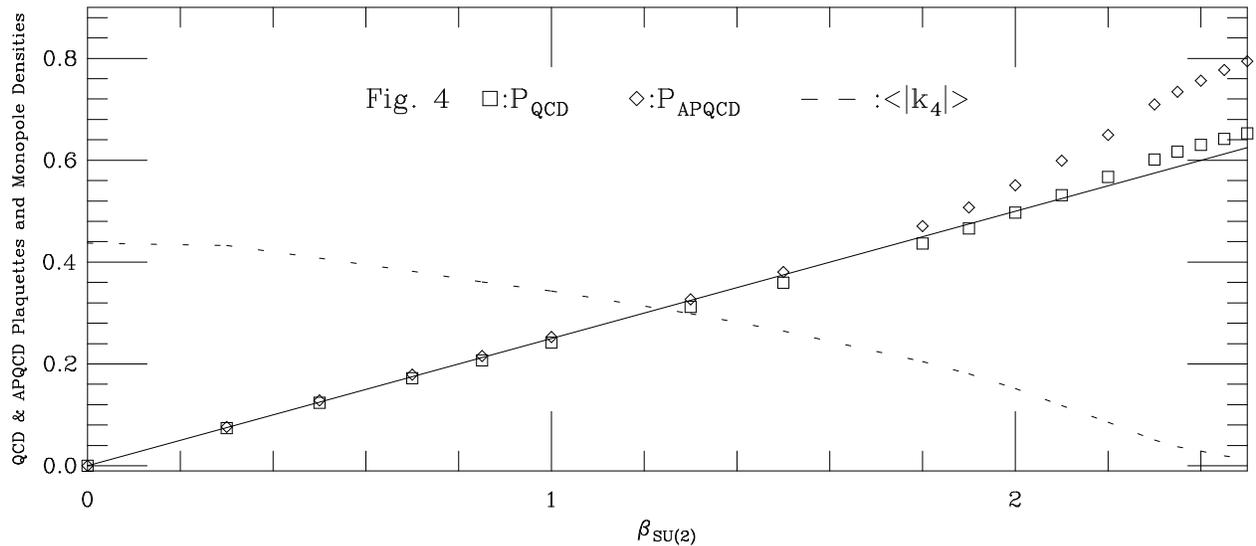

Fig. 4. Figure 4 depicts the APQCD and $SU(2)$ plaquettes as a function of $\beta_{SU(2)}$. In the strong coupling region($\beta_{SU(2)} < 2$), both the APQCD and $SU(2)$ plaquettes grow like $\frac{1}{4}\beta_{SU(2)}$, the guide-to-eye line's slope. At weaker coupling($\beta_{SU(2)} > 2$) the APQCD plaquettes deviate substantially from the $SU(2)$ plaquettes. Correspondingly, the monopole density decelerates noticeably near $\beta_{SU(2)} \sim 2$.

to determine if these lattice results have a true continuum significance.

## 4. Acknowledgements

It is a pleasure to thank Misha Polikarpov for many stimulating discussions and for the use of his $SU(2)$ FORTRAN codes. I am indebted to the faculty and students of the Institute for Theoretical and Experimental Physics(ITEP) for their generous hospitality. Computing was done at the NERSC Supercomputer Center. The author is supported by DOE grant DE-FG05-91ER40617.